\documentclass[12pt]{article}
\usepackage{amsmath}
\usepackage{amssymb}

\usepackage{scicite}

\usepackage{subfigure}
\usepackage{times}
\usepackage{epsfig}
\usepackage{url}

\newenvironment{sciabstract}{%
\begin{quote} \bf}
{\end{quote}}

\newcounter{lastnote}

\begin{document}


\title{ Community Structures Are Definable in Networks, and Universal in Real World%
\footnote{The research is fully supported by the Grand Project
``Network Algorithms and Digital Information'' of the Institute of
Software, Chinese Academy of Sciences, and partially supported by
the NSFC grant number 61161130530, and by a 973 program of grant
number 2014CB340302.}}

\author{ Angsheng Li$^{1}$,  Jiankou Li$^{1,2 }$, Yicheng Pan$^{1,3}$ \\
\normalsize{$^{1}$State Key Laboratory of Computer Science}\\
\normalsize{ Institute of Software, Chinese Academy of Sciences}\\
\normalsize{$^{2}$University of Chinese Academy of Sciences,
P. R. China}\\
\normalsize{$^{3}$State Key Laboratory of Information Security}\\
\normalsize{ Institute of Information Engineering, Chinese Academy of Sciences,
P. R. China} }


\date{}



\baselineskip24pt

\maketitle

\begin{sciabstract}

  Community detecting is one of the main approaches to understanding
   networks \cite{For2010}.
  However it has been a longstanding
   challenge to give a definition for community structures of networks. Here we
found that community structures are definable in networks, and are
universal in real world. We proposed the notions of entropy- and
conductance-community structure ratios. It was shown that the
definitions of the modularity proposed in \cite{NG2004}, and our entropy- and conductance-community
structures are equivalent in defining community structures of
networks, that randomness in
the ER model \cite{ER1960} and preferential attachment in the PA
\cite{Bar1999} model are not mechanisms of community structures of
networks, and that the existence of community structures is a
universal phenomenon in real networks. Our results demonstrate that
community structure is a universal phenomenon in the real world that is definable,
 solving the challenge of definition of community structures in networks. This progress provides a
foundation for a structural theory of networks.

\end{sciabstract}

We proposed a definition of community structures in networks, solving the fundamental challenge
in modern network theory. Our definitions of the entropy- and
conductance-community structures are information theoretical and
mathematical definitions respectively. Our result of the equivalence of our
entropy-, and conductance-community structure ratios, together with the modularity given by physicists
shows that the existence of community structures in networks is a phenomenon definable
by each of the physical, information theoretical and mathematical approaches, providing a common
foundation for the interdisciplinary issue of networks. Our definitions
of community structures of networks provide a method to decide both the existence and the quality
of community structures of networks. Our discovery of the universality of community structures
of real networks predicts that community structures maybe universal in the real world data, and that
community structures maybe the key to a structural theory of networks and real world data in general.
Our discovery that neither randomness nor preferential attachment is the mechanism of community structures of networks
predicts that there must be new mechanisms for real world data. Therefore the definitions and discoveries here
not only provide a foundation for a new theory of networks, but also a methodology
for rigorous analysis of real world data.

{\bf Results}

Network  has become a universal topology in science, industry,
nature and society. Most real networks satisfy a power law degree
distribution~\cite{Bar1999, Bar2009}, and a small world
phenomenon~\cite{M1967, W1998, K2000}.

Community detecting or clustering is a powerful tool for
understanding the structures of networks, and has been extensively
studied ~\cite{EK2010, CR2010, CA2005, RCCLP2004, CNM2004, NEW2004}.
Many definitions of communities have been introduced, see
\cite{For2010} for a recent survey. However, the problem is still
very hard, not yet satisfactorily solved. The current approaches to community
finding take for granted that networks have community structures.
The fundamental questions are thus: Are communities objects
naturally formed in a network or simply outputs of a graphic
algorithm? Can we really take for granted that networks have
community structures? Are community structures definable in
networks? What are the natural mechanisms of the community structure
of a network, if any?

Here we report our discovery that community structures are robust in
networks, in the sense that, the three definitions of community
structures based on modularity, entropy and conductance respectively
give the same answer to the question whether or not a network has a
community structure, that community structures are universal in real
networks, and that neither randomness nor preferential attachment is
the mechanism of community structures of networks.

{\bf Modularity, Entropy and Conductance Definitions of Community
Structure}

The first definition is the modularity community structure
(M-community structure, for short). Newman and Girvan \cite{NG2004}
defined the notion of modularity to quantitatively measure the
quality of community structure of a network. It is built based on
the assumptions that random graphs are not expected to have
community structure and that a network has a community structure, if
it is far from random graphs.

Let $G=(V,E)$ be a network. Given a partition $\mathcal{P}$ of $G$,
the modularity of the partition $\mathcal{P}$ of network $G$ with
$n$ nodes and $m$ edges is defined by

$$q^{\mathcal{P}}(G)=\frac{1}{2m} \sum\limits_{i,j} (A_{ij}-P_{ij})\delta (C_i,
C_j),$$

\noindent where the summation runs over all pairs of vertices, $A$
is the adjacency matrix, $P_{ij}$ is the expected number of edges
between vertices $i$ and $j$ in a null graph, i.e., a random version
of $G$. $\delta (C_j,C_j)=1$ if $C_i=C_j$, and $0$ otherwise, $C_k$
is an element of the partition $\mathcal{P}$.

The modularity of $G$ is defined by

$$\sigma (G)=\max\limits_{\mathcal{P}}\{q^{\mathcal{P}}(G)\}.$$

Intuitively speaking, the larger $\sigma (G)$ is, the better
community structure $G$ has. Therefore we define the {\it modularity
community structure ratio } (M-community structure ratio) of $G$ to
be the modularity of $G$.

The second definition is based on random walks. The idea is that
since random walks from a node in a quality community are not easy
to go out of the community, a network can be decomposed into
modules by compressing the description of an information flow. Rosvall and Bergstrom \cite{RB2008}
proposed a way to use the Huffman code to encode prefix-freely each module and each
node (adding an exit code) of a network. This allows us to reuse the codeword of a module-node
for a random walk within the module, which compresses the bits of descriptions of random walks
by the modules, compared to that of a uniform prefix-free code for all nodes.

Our definition follows the same idea. We consider the shortest average length of codes for a single step of random
walks in the case of the standard stationary distribution that the
probability of staying at some node $i$ is proportional to the
degree of $i$.

Let $G=(V,E)$ be a graph with $n$ nodes and $m$ edges, and
$\mathcal{P}$ be a partition of $V$. We use $L^U(G)$ to denote the
minimum average number of bits to represent a step of random walk
(in the stationary distribution) with a uniform code in $G$, and
$L^{\mathcal{P}}(G)$ to denote the minimum average number of bits to
represent a step of random walk in $G$ with a code of modules given by $\mathcal{P}$ in $G$. By
information theoretical principle, we have
\begin{equation} \label{eqn:entropy_G}
L^U(G)=-\sum\limits_{i=1}^n \frac{d_i}{2m} \cdot
\log_2\frac{d_i}{2m},
\end{equation}
where $d_i$ is the degree of node $i$.

\begin{equation} \label{eqn:entropy_partition}
L^{\mathcal{P}}(G)=-\sum\limits_{j=1}^L \sum\limits_{i=1}^{n_j}
\frac{d_i^{(j)}}{2m} \cdot \log_2\frac{d_i^{(j)}}{V_j}-\frac{m_g}{m}
\left( \sum\limits_{j=1}^L \frac{V_j}{2m} \cdot \log_2\frac{V_j}{2m}
\right),
\end{equation}
where $L$ is the number of modules in partition $\mathcal{P}$, $n_j$
is the number of nodes in module $j$, $d_i^{(j)}$ is the degree of
node $i$ in module $j$, $V_j$ is the volume of module $j$, and $m_g$
is the number of edges crossing two different modules.

We define the {\it entropy community structure ratio of $G$ by
$\mathcal{P}$} by

$$\tau^{\mathcal{P}} (G) = 1 - \frac{L^{\mathcal{P}}(G)}{ L^U(G)}.$$

We define the {\it entropy community structure ratio of $G$}
(E-community structure ratio of $G$) by

$$\tau (G)=\max_{\mathcal{P}}\{\tau^{\mathcal{P}}(G)\}.$$

Both the modularity and the entropy community structure ratio of a
graph $G$ depend on randomness, the first is in the null version of
the graph, and the second is in random walks in the graph. The two definitions are
not convenient to measure the quality of overlapping communities, instead of a partition
of the graph.

Here we introduce a mathematical definition based on conductance. Given a
graph $G=(V,E)$, and a subset $S$ of $V$, the conductance of $S$ is
given by

$$\Phi (S)=\frac{|E(S,\bar{S})|}{\min\{ {\rm vol}(S), {\rm vol
}(\bar{S})\}},$$

\noindent where $E(S,\bar{S})$ is the set of edges with one endpoint
in $S$ and the other in the complement of $S$, i.e. $\bar{S}$, ${\rm
vol}(X)$ is the summation of degrees $d_x$ for all $x\in X$.

We say that a set $X\subset V$ is  a {\it possible community} if:
(i) the induced subgraph of $X$, $G_X$ is connected, (ii) the size
$|X|$ of $X$ is not less than $\log n$ (i.e., not too small), and
(iii) the size of $X$ is less than $\sqrt{n}$ (i.e., not too large),
where $n$ is the size of $V$.

(i) is a basic condition. (ii) and (iii) avoid trivial communities
that are either not well-evolved, or is essentially a significant
part of the whole network.

Suppose that $\mathcal{X}=\{X_1, X_2,\cdots, X_l\}$ is a set of
 possible communities of $G$. Let $X=\cup_jX_j$. For a community $X_j$, we use $1-\Phi (X_j)$
to define the quality of the community.

For every $x\in X$, suppose that $X_1', X_2', \cdots, X_p'$ are all
$X_j$'s that contain $x$, then define

$$a^{\mathcal{X}}(x)=\frac{1}{p}\sum\limits_{j=1}^p (1-\Phi (X_j')),$$

\noindent where $a^{\mathcal{X}}(x)$ represents the average quality
of all the communities containing $x$.

We define  the {\it conductance community structure ratio of $G$ by
$\mathcal{X}$} (or C-community structure ratio, for short) by

$$\theta^{\mathcal{X}}(G)=\frac{1}{n}\sum\limits_{x\in X} a^{\mathcal{X}}(x),$$

\noindent where $n$ is the number of nodes in $G$.

We define the {\it conductance community structure ratio of $G$} by

$$\theta (G)=\max_{\mathcal{X}}\{\theta^{\mathcal{X}}(G)\}.$$

Let $\mathcal{A}$ be an algorithm, and $G$ be a network. Suppose
that $\mathcal{X}$ is the set of all possible communities found in $G$ by
$\mathcal{A}$. Then define the {\it conductance community structure
ratio of $G$ by $\mathcal{A}$} by

$$\theta^{\mathcal{A}}(G)=\theta^{\mathcal{X}}(G).$$

 This gives rise
to a way to measure the quality of a community detecting algorithm.
Intuitively, for two algorithms $\mathcal{A}$ and $\mathcal{B}$, if
$\theta^{\mathcal{A}}(G)>\theta^{\mathcal{B}}(G)$, then
$\mathcal{A}$ is better than $\mathcal{B}$ in finding the community
structure of $G$. Clearly $\theta (G)$ characterizes the community structure of $G$.

Now we have three definitions of community structure of networks,
the M-, E-, and C-community structure ratios. Intuitively speaking, the M-, E- and
C-community structure ratios capture the quality of
community structure of $G$ from the viewpoints of physics,
information theory and mathematics respectively.

{\bf The Modularity, Entropy and Conductance Definitions of
Community Structure Are Equivalent}

Are there any relationships among the three definitions of quality
of community structures of networks, i.e., the M-, E-, and
C-community structure ratios? Do the three definitions give the same
answer to the question whether or not a network has a community
structure? We conjecture that the answer is yes. For this, we
propose the following hypothesis.

 {\it Community structure hypothesis}: Given a network $G$, the following properties are
equivalent,

\begin{enumerate}

\item [1)] $G$ has an M-community structure,

\item [2)] $G$ has an E-community structure, and

\item [3)] $G$ has a C-community structure.

\end{enumerate}

We verify the community structure hypothesis by computing the M-,
E-, and C-community structure ratios for networks of classical
models. The first model is the ER model \cite{ER1960}. In this
model, we construct graph as follows: Given $n$ nodes $1,2,\cdots,
n$, and a number $p$, for any pair $i, j$ of nodes $i$ and $j$, we
create an edge $(i,j)$ with probability $p$. The second is the PA
model \cite{Bar1999}. In this model, we construct a network by steps
as follows: At step $0$, choose an initial graph $G_0$. At step
$t>0$, we create a new node, $v$ say, and create $d$ edges from $v$
to nodes in $G_{t-1}$, chosen with probability proportional to the
degrees in $G_{t-1}$, where $G_{t-1}$ is the graph constructed at
the end of step $t-1$, and $d$ is a natural number.

\begin{figure}
  \centering
  \includegraphics[width=4in]{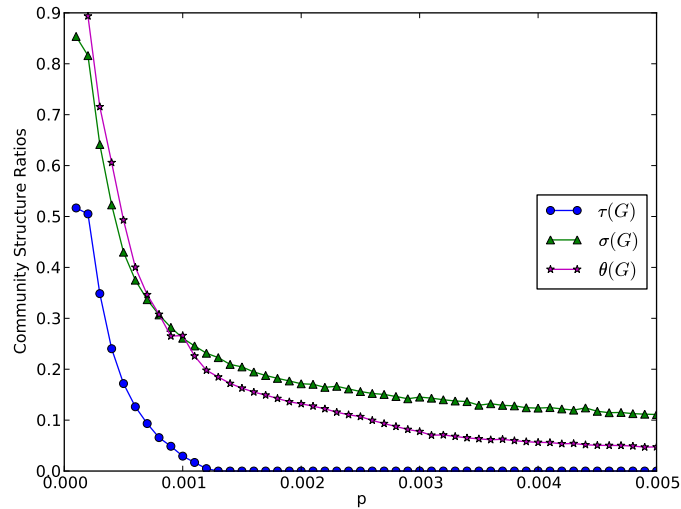}
  \caption{{This figure gives the E-, M- and C-community structure ratios
  (denoted by e-, m- and c-ratios respectively) of networks, for $n=10,000$, and for $p$ up to $0.005$ of the ER model.}}
\label{figure_er_modularity}
\end{figure}

\begin{figure}
  \centering
  \includegraphics[width=4in]{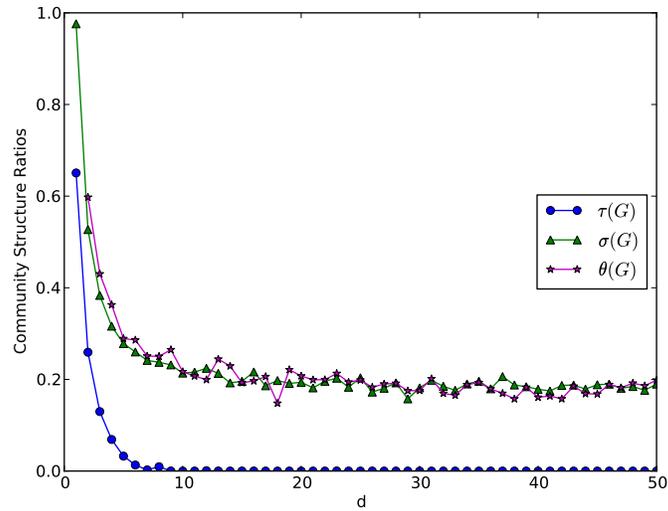}
  \caption{{This figure gives the E-, M- and C-community structure ratios
  (denoted by e-, m- and c-ratios respectively) of networks, for $n=10,000$, and for $d\leq 50$ of the PA model.}}\label{figure_ba_modularity}
\end{figure}

We depict the curves of the M-, E-, and C-community structure ratios
of networks of the ER model and the PA model in Figures
~\ref{figure_er_modularity} and ~\ref{figure_ba_modularity}
respectively.

From Figures \ref{figure_er_modularity} and
\ref{figure_ba_modularity}, we observe that:

\begin{enumerate}
\item [(1)] The curves of the M-, E-, and C-community structure ratios of
networks generated from the ER model are similar.
\item [(2)] The curves of the M-, E-, and C-community structure
ratios of networks generated from the PA model are similar.

\end{enumerate}

(1) and (2) show that the community structure hypothesis holds for
all networks generated from the classic ER and PA models. We notice
that every network essentially uses the mechanisms of both the ER
and the PA models. Our results here imply that the community
structure hypothesis may hold for most real networks.

{\bf Empirical Criterions of Community Structures}

By observing the experiments in Figures~\ref{figure_er_modularity}
and ~\ref{figure_ba_modularity}, we have that for a network $G$ of
either the ER model or the PA model, the following three properties
(1), (2) and (3) either  hold simultaneously or fail to hold
simultaneously:

\begin{enumerate}
\item [(1)] the E-community structure ratio of $G$, $\tau (G)$, is
greater than $0$,

\item [(2)] the M-community structure ratio of $G$, $\sigma (G)$, is
greater than $0.3$, and

\item [(3)] the C-community structure ratio of $G$, $\theta (G)$, is
greater than $0.3$.
\end{enumerate}

This result suggests an {\it empirical criterion} for deciding
whether or not a network has a community structure. Let $G$ be a network, then

\begin{enumerate}
\item We say that $G$ has a community structure if the E-, M-, and C-community structure ratios of $G$
are greater than $0$, $0.3$ and $0.3$ respectively.

\item The values $\sigma (G)$, $\tau (G)$ and $\theta (G)$ measure
the quality of community structure of $G$, the larger they are, the
better community structure $G$ has.

\end{enumerate}

{\bf Randomness and Preferential Attachment Are Not Mechanisms of
Community Structure}

By the empirical criterion and by observing the experiments in
Figures~\ref{figure_er_modularity} and ~\ref{figure_ba_modularity},
we have that

\begin{enumerate}
\item For a network $G$ generated from the ER model, if $p<\frac{1}{2000}$ (in which case, the expected average number of edges is $<5$), then
$G$ has a community structure, and if $p>\frac{1}{2000}$, then $G$
fails to have a community structure.

\item For a network $G$ generated from the PA model, if $d<5$, then
$G$ has a community structure, and if $d>5$, then $G$ fails to have
a community structure.

\end{enumerate}

This shows that the existence of community structure of networks of
the ER and PA models depends on the density of the networks, that
only networks with average number of edges $<5$ may have a community
structure, and that nontrivial networks of the ER and PA models fail
to have a community structure. This is an interesting and useful
discovery. It explains some mysterious phenomena: usually people
believe that networks generated from the ER and PA models fail to
have a community structure (although a proof is apparently needed),
but sometimes people found graphs of the ER and PA models having
extremely high modularity \cite{For2007}; in evolutionary games,
some people implemented experiments on networks of the PA model with
particular average number of edges $d=4$ without any
explanation~\cite{SP2006, SPL2006}. Now we know that a network of
the ER or PA model has a community structure only if the average
number of edges is less than a small constant, $5$ say, and that
community structure of a network plays an essential role in
networks.

{\bf Community Structures Are Universal in Real Networks}

By using the empirical criterion of community structure of networks,
we are able to decide whether or not a given network has a
community structure.

We implemented the experiments of the entropy-, modularity- and
conductance-community structure ratios, i.e., $\tau (G)$, $\sigma
(G)$ and $\theta (G)$, for $22$ real networks, which are given in
Table~\ref{table_statistic}. By observing the table, we have the
following results: For every network $G$,

\begin{enumerate}

\item [(1)] Then:

-- $\tau (G)>0$,

-- $\sigma (G)>0.3$, and

-- $\theta (G)>0.3$.

\item [(2)] $\tau (G)\leq \sigma (G)$ and $\tau (G)\leq\theta (G)$.

\item [(3)] For most networks $G$, $\sigma (G)\approx\tau (G)+\alpha$ for some number $\alpha$ in the
interval $[0.2,0.3]$, and $\sigma (G)\approx\theta (G)$.

\end{enumerate}

The experiments in Table~\ref{table_statistic} show that the
community structure hypothesis holds for real networks, that
community structures are universal in most real networks, and that
the existence of community structures in real networks is
independent of which definition of the M-, E- and C-community
structures is used.

By observing all the curves in Figures ~\ref{figure_er_modularity}
and ~\ref{figure_ba_modularity}, and all experiments in Table
\ref{table_statistic} again, we have the following conclusions:
(1) The three definitions of modularity-, entropy- and conductance-community
structures are equivalent in defining community structures of
networks. This implies that the physical, information theoretical, and
mathematical definitions of community structures of networks are
equivalent, and that the existence of community structures of
networks is a phenomenon independent of which one of the physical,
information theoretical and mathematical definitions of community
structures is used, and independent of algorithms for finding them.
(2) There exists an empirical criterion for deciding the existence and quality
of community structure of a network. This also solves an important open question to
test the quality of community finding algorithms.
(3) Neither randomness nor preferential attachment is a mechanism
of community structures of networks.
(4) Community structures are universal in real networks.
Together with (1) above, this implies that the existence of
community structures is a universal phenomenon of real networks, for
which we have to explain the reason why. Together with  (3) above, this implies that
there must be new mechanisms for the existence of community
structures of real networks other than the well-known mechanisms of
randomness and preferential attachment for classic models of
networks.

{\bf Discussions}

Our results above show that the physical, information theoretical
and mathematical definitions of community structures of networks are
equivalent in characterizing the existence and quality of community
structures of networks, that nontrivial networks of classic ER and
PA models fail to have a community structure, and that most real
networks do have a community structure. The significance of our results are four folds: 1) the existence of
community structures is a natural phenomenon definable in networks, by one of the physical, information theoretical and mathematical definitions,
2) community structures are universal in real world data, 3) mechanisms of classic models are not mechanisms of community structures of networks,
and 4) the existence and quality of community structures of network data can be tested by our definitions and criterions. This progress
poses fundamental questions: What are the
mechanisms of community structures of real networks? What roles do
the community structures play in networks? What are the new algorithms and applications based on structures of networks and big data, in general?
Answering these questions
would build a new theory of networks, the structural theory of networks,
which is of course a grand challenge in network science.

{\bf Methods }

The data of real networks can be found from the websites:
\url{http://snap.standford.edu}, or
\url{http://www-personal.umich.edu/~mejn/netdata}.

\begin{table}
 \centering
\begin{tabular} {|c|c|c|c|}
\hline
network &$\tau (G)$&$\sigma (G)$&$\theta (G)$\\
\hline
cit\_hepph&0.22&0.56&0.37\\
\hline
cit\_hepth&0.2&0.53&0.36\\
\hline
col\_astroph&0.24&0.51&0.49\\
\hline
col\_condmat&0.37&0.64&0.76\\
\hline
col\_grqc&0.44&0.79&0.89\\
\hline
col\_hepph&0.26&0.58&0.7\\
\hline
col\_hepth&0.39&0.69&0.83\\
\hline
email\_enron&0.21&0.5&0.63\\
\hline
email\_euall&0.39&0.73&0.76\\
\hline
p2p4&0.11&0.38&0.36\\
\hline
p2p5&0.11&0.4&0.36\\
\hline
p2p6&0.12&0.39&0.38\\
\hline
p2p8&0.15&0.46&0.46\\
\hline
p2p9&0.15&0.46&0.42\\
\hline
p2p24&0.21&0.47&0.48\\
\hline
p2p25&0.23&0.49&0.5\\
\hline
p2p30&0.24&0.5&0.53\\
\hline
p2p31&0.25&0.5&0.52\\
\hline
roadnet\_ca&0.67&0.99&0.98\\
\hline
roadnet\_pa&0.66&0.99&0.98\\
\hline
roadnet\_tx&0.67&0.99&0.98\\
\hline
\end{tabular}
 \caption{The entropy, modularity and
conductance community structure ratios of real networks, written by
$\tau (G)$, $\sigma (G)$ and $\theta (G)$ respectively.}
\label{table_statistic}
\end{table}


{\bf Acknowledgements}

All authors are partially supported by the Grand Project ``Network
Algorithms and Digital Information'' of the Institute of Software,
Chinese Academy of Sciences, by an NSFC grant No. 61161130530 and a
973 program grant No. 2014CB340302. The third author is partially supported by a National Key Basic Research Project of China (2011CB302400)
and  the "Strategic Priority Research Program" of the Chinese Academy of
Sciences£¬Grant No. XDA06010701.

{\bf Author Contributions} AL designed the research and wrote the
paper, JL and YP performed the research.

{\bf Additional information }

Competing financial interests: The authors declare they have no
competing financial interests.

\end{document}